\documentclass[aps,showpacs,preprintnumbers,amsmath,amssymb,twocolumn]{revtex4}
\usepackage{graphicx}

\begin{document}

\title{On the possibility of hyperbolic tunneling}

\author{Francisco S. N. Lobo}%
\email{flobo@cii.fc.ul.pt} \affiliation{Centro de F\'{i}sica
Te\'{o}rica e Computacional, Faculdade de Ci\^{e}ncias da
Universidade de Lisboa, Campo Grande, Ed. C8 1749-016 Lisboa,
Portugal}

\author{Jos\'{e} P. Mimoso}
\email{jpmimoso@cii.fc.ul.pt} \affiliation{Centro de F\'{i}sica
Te\'{o}rica e Computacional, Faculdade de Ci\^{e}ncias da
Universidade de Lisboa, Campo Grande, Ed. C8 1749-016 Lisboa,
Portugal}

\date{\today}

\begin{abstract}

Traversable wormhole are primarily useful as
``gedanken-experiments'' and as a theoretician's probe of the
foundations of general relativity. In this work, we analyse the
possibility of having tunnels in a hyperbolic spacetime. We obtain
exact solutions of static and {\it pseudo}-spherically symmetric
spacetime tunnels by adding exotic matter to a vacuum solution
referred to as a degenerate solution of class $A$. The physical
properties and characteristics of these intriguing solutions are
explored, and through the mathematics of embedding it is shown
that particular constraints are placed on the shape function, that
differ significantly from the Morris-Thorne wormhole. In
particular, it is shown that the energy density is always negative
and the radial pressure is positive, at the throat, contrary to
the Morris-Thorne counterpart. Specific solutions are also
presented by considering several equations of state, and by
imposing restricted choices for the shape function or the redshift
function.

\end{abstract}


\maketitle


\section{Introduction}

Wormholes are hypothetical tunnels in spacetime, possibly through
which observers may freely traverse. However, it is important to
emphasize that these solutions are primarily useful as
``gedanken-experiments'' and as a theoretician's probe of the
foundations of general relativity (GR). In classical general
relativity, wormholes are supported by exotic matter, which
involves a stress-energy tensor that violates the null energy
condition (NEC) \cite{Morris:1988cz,Visser}. Note that the NEC is
given by $T_{\mu\nu}k^\mu k^\nu \geq 0$, where $k^\mu$ is {\it
any} null vector. Thus, it is an important and intriguing
challenge in wormhole physics to find a realistic matter source
that will support these exotic spacetimes. Several candidates have
been proposed in the literature, amongst which we refer to
solutions in higher dimensions, for instance in
Einstein-Gauss-Bonnet theory \cite{EGB1,EGB2}, wormholes on the
brane \cite{braneWH1,braneWH2}; solutions in Brans-Dicke theory
\cite{Nandi:1997en}; wormhole solutions in semi-classical gravity
(see Ref. \cite{Garattini:2007ff} and references therein); exact
wormhole solutions using a more systematic geometric approach were
found \cite{Boehmer:2007rm}; geometries supported by equations of
state responsible for the cosmic acceleration
\cite{phantomWH,ChapWH,vdWWH}, and solutions in conformal Weyl
gravity were also found \cite{Weylgrav}, etc (see Refs.
\cite{Lemos:2003jb,Lobo:2007zb} for more details and
\cite{Lobo:2007zb} for a recent review).

In this work, instead of further exploring wormholes in some
extension of GR, we rather analyse the possibility of having
tunneling solutions in general relativistic solutions with
negatively curved spatial surfaces which might be considered the
analogue of wormholes. Indeed, we consider a largely ignored
metric which belongs to a class of vacuum solutions referred as
degenerate solutions of class A class A \cite{Stephani:2003tm} by
Ehlers and Kundt~\cite{Ehlers & Kundt 1962,Levi-Civita 1917a},
which are axisymmetric solutions and thus also belong to Weyl's
class \cite{Weyl:1917gp}, given by
\begin{equation}
ds^2= -e^{\mu(r)}\,{\rm d}t^2+ e^{\lambda(r)} \,{\rm
d}r^2+r^2\,({\rm d}u^2+\sinh^2{u}\,{\rm d}v^2) \,,
\label{metric_constnc}
\end{equation}
where the usual 2$-d$ spheres are replaced by pseudo-spheres,
${\rm d}\sigma^2={\rm d}u^2+\sinh^2{u}\,{\rm d}v^2$, hence by
surfaces of negative, constant curvature. These are still surfaces
of revolution around an axis, and $v$ represents the corresponding
rotation angle. The specific case of
\begin{equation}
e^{\mu(r)}=e^{-\lambda(r)} =\left(\frac{2\mu}{r}-1\right) \; ,
\label{metric_constnc-antiScwarz}
\end{equation}
is of particular interest, where $\mu$ is a constant \cite{Stephani:2003tm,Bonnor
& Martins 1991,Martins 1996}.

In our opinion this metric can be seen as an anti-Schwarzschild in
the same way the de Sitter model with negative curvature is an
anti-de Sitter model. We immediately see that the static solution
holds for $r<2\mu$ and that there is a coordinate singularity at
$r=2\mu$ (note that $|g|$ neither vanishes nor becomes $\infty$ at
$r=2\mu$)\cite{Anchordoqui:1995wa}. This is the complementary
domain of the exterior Schwarzschild solution in the region
outside the Schwarzchild horizon. In the domain $r>2\mu$, as with
the latter solution, the $g_{tt}$ and $g_{rr}$ metric coefficients
swap signs. Defining $\tau=r$ and $\rho=t$, we obtain
${\rm d} s^2= -{\rm d}\tilde\tau^2+A^2(\tilde{\tau})\,{\rm
d}\rho^2 +B^2(\tilde\tau)\,({\rm d}u^2+\sinh^2{u}\,{\rm d}v^2),$
with the following parametric definitions $\tilde{\tau} = -\tau
+2\mu \, \ln|\tau-2\mu|$, $A^2=2\mu/\tau-1$ and
$B^2(\tau)=\tau^2$, which is a particular case of a Bianchi III
axisymetric universe.

Using pseudo-spherical coordinates $\{x = r\sinh u  \cos v,\, y =
r\sinh u  \sin v,\,z = r \cosh u,\,w = b(r)\}$, the spatial part
of the metric (\ref{metric_constnc}) can be related to the
hyperboloid $w^2+x^2+y^2-z^2 = \left(b^2/r^2 - 1\right)\, r^2$
embedded in a 4-dimensional flat space. We then have
\begin{eqnarray}
{\rm d}w^2 +{\rm d}x^2+{\rm d}y^2-{\rm d}z^2 &= & \left[(b'(r))^2 -
1\right]\,{\rm d}r^2+ \nonumber \\ && r^2 \left( {\rm d}u^2+\sinh^2u\, {\rm d}v^2
\right) \; .
\end{eqnarray}
where the prime stands for differentiation with respect to $r$,
and $b(r) = \mp 2\sqrt{2\mu}\sqrt{2\mu-r}$. We can recast metric
(\ref{metric_constnc}) into the following
\begin{eqnarray}
{\rm d} s^2= & -\, \tan^2\left[\ln\left( \bar r\right)^{\mp 1}
\right] \,{\rm d}\tau^2+
 \left(\frac{2\mu}{\bar r}\right)^2\,
 \cos ^4\left[\ln\left( \bar r\right)^{\mp 1}\right]\times\nonumber \\ & \times\, \left[{\rm d}\bar{r}^2\,+ \bar r^2\,({\rm d}u^2
+\sinh^2{u}\,{\rm d}v^2) \right]\; , \label{Isotrop3}
\end{eqnarray}
which is the analogue of the isotropic form of the Schwarzschild
solution. In the neighborhood of $u=0$, i.e., for $u\ll 1$, we can
cast the metric of the 2-dimensional hyperbolic solid angle as
\begin{equation}
{\rm d}\sigma^2 \simeq  {\rm d}u^2 + u^2 {\rm d}v^2 \,
\end{equation}
so that it confounds itself with the tangent space to the
spherically symmetric $S^2$ surfaces in neighborhood of the poles.
The apparent arbitrariness of the locus $u=0$, is overcome simply
by transforming it to another location by means of a hyperbolic
rotation, as in the case of the spherically symmetries case where
the poles are defined up to a spherical rotation (SO(3) group).
Thus, the spatial surfaces are conformally flat. However, we
cannot recover the usual Newtonian weak-field limit for large $r$,
because of the change of signature that takes place at $r=2\mu$.

Analyzing the ``radial'' motion of test particles, we have the
following equation
\begin{equation}
\dot{r}^2+ \left(\frac{2\mu}{r} -1\right)\,\left(1+\frac{h^2} {r^2
\sinh^2 u_\ast } \right) = \epsilon
\end{equation}
where $\epsilon$ and $h$ are constants of motion defined by
$\epsilon= \left(2\mu/r -1\right)\, \dot{t} = {\rm const_t}$ and
$h^2= r^2 \,\sinh^2 u_\ast \, \dot{v} = {\rm const_v}$, for fixed
$u=u_\ast$. The former and latter constants represent the energy
and angular momentum per unit mass, respectively. We thus define
the potential
$2V(r) = \left(\frac{2\mu}{r} -1\right)\,\left(1+\frac{h^2} {r^2
\sinh^2 u_\ast } \right).\label{geod_potential}$
This potential is manifestly repulsive, crosses the $r$-axis at
$r=2\mu$, and for sufficiently high values of $h$ it has a minimum
at
$r_{\pm} =(h^2\mp \sqrt{h^4 - 12 \mu^2 h^2})/(2\mu)$.
However this minimum 
falls outside the $r=2\mu$ divide. So a test particle is subject
to a repulsive potential forcing it to inevitably cross the event
horizon at $r=2\mu$ attracted either by some mass at the minimum
or by masses at infinity. In \cite{Bonnor & Martins 1991} it is
hinted that the non-existence of a clear Newtonian analogue is
related to the existence of mass sources at $\infty$, but no
definite conclusions were drawn \footnote{The puzzling features of
this metric are, seemingly, the reason why it was coined
degenerate by Ehlers and Kundt~\cite{Ehlers & Kundt 1962}.}.

A more detailed analysis of the physical properties and
characteristics of this intriguing solution is presently underway
\cite{MimLobo1}, as well as an application to the cosmological
features of the model, namely the study of negatively curved
spacetimes in order to understand the ultimate stages of
underdensities \cite{MimLobo2}.

Here we shall study the extension of the solution
(\ref{metric_constnc-antiScwarz}) which arises from adding exotic
matter to analyze the possibility of tunneling in hyperbolic
spacetimes. These would add to other non-spherically symmetric
wormholes which have already been considered in the literature.
For instance, extending the spherically symmetric Morris-Thorne
wormholes \cite{Morris:1988cz} and motivated by the aim of
minimizing the violation of the energy conditions, polyhedral
solutions and, in particular, cubic wormholes were constructed in
Ref. \cite{Visser89}. In Ref. \cite{GDiaz}, the static spherically
symmetric traversable wormhole solution was generalized to that of
a (non-planar) torus-like topology \cite{GDiaz}, denoted as a
ringhole. In Ref. \cite{LL}, solutions of plane symmetric
wormholes in the presence of a negative cosmological constant by
matching an interior spacetime to the exterior anti-de Sitter
vacuum solution were constructed. It is interesting to note that
the construction of these plane symmetric wormholes does not alter
the topology of the background spacetime (i.e., spacetime is not
multiply-connected), so that these solutions can instead be
considered domain walls. The dynamic stability analysis of plane
symmetric wormholes was further analyzed in Ref. \cite{LL2}.

Thus, it is the purpose of this paper to study static and {\it
pseudo}-spherically symmetric counterparts to the usual wormholes
by adding exotic matter to the vacuum degenerate solution of class
$A$, given by (\ref{metric_constnc}).
The physical properties and characteristics of these intriguing
solutions are explored, and through the mathematics of embedding
it is shown that particular constraints are placed on the shape
function, that differ radically from the Morris-Thorne wormhole.
In particular, it is shown that the energy density is always
negative and the radial pressure is positive, at the throat,
contrary to the Morris-Thorne counterpart. Specific solutions are
also presented by considering various equations of state, and by
imposing restricted choices for the shape function or the redshift
function.

This paper is organized in the following manner: In Sec.
\ref{Sec:II}, the spacetime metric, the field equations and the
mathematics of embedding of these pseudo-spherically symmetric
geometries are analyzed in detail. In Sec. \ref{Sec:III}, specific
solutions are found by considering several equations of state. We
conclude in Sec. \ref{Sec:conclusion}.


\section{Traversable tunnels in pseudo-spherical symmetry}
\label{Sec:II}

\subsection{Spacetime metric and field equations}

Consider the following {\it pseudo}-spherically symmetric and
static  solution
\begin{equation}
{\rm d}s^2=-e ^{2\Phi(r)} \,{\rm d}t^2+\frac{{\rm
d}r^2}{b(r)/r-1}+r^2 \,({\rm d}u ^2+\sinh ^2{u} \, {\rm d}v ^2)
\,, \label{metricwormhole}
\end{equation}
where the coordinates $u$ and $v$ have the following range
$-\infty<u<+\infty$ and $0\leq v \leq 2\pi$. The $r$ coordinate
yields the curvature radius of the 2-dimensional pseudo-spherical
surfaces that thread the spacetime: $^{(2)}R=- r^{-2}$, and,
hence, is a generalised radial coordinate. $\Phi(r)$ and $b(r)$
are arbitrary functions of the radial coordinate $r$. As in the
Morris-Thorne wormhole \cite{Morris:1988cz}, we denote $\Phi(r)$
the redshift function, for it is related to the gravitational
redshift, and $b(r)$ the shape function , as will be shown below
by embedding diagrams, determines the shape of the tunnel (in
analogy to the analysis considered in \cite{Morris:1988cz}). The
coordinate $r$ is non-monotonic in that it decreases from a
constant value $C$ to a minimum value $r_0$, representing the
location of the throat of the wormhole, where $b(r_0)=r_0$, and
then it increases from $r_0$ back to the value $C$. Note that the
condition $(b/r-1) \geq 0$ imposes that $b(r)\geq r$, contrary to
the Morris-Thorne counterpart.

The Einstein field equation, $G_{\mu\nu}=8\pi T_{\mu\nu}$,
provides the following stress-energy scenario
\begin{eqnarray}
\rho(r)&=&-\frac{1}{8\pi} \, \frac{b'}{r^2}
\label{rhoWH}\,,\\
p_r(r)&=&\frac{1}{8\pi} \, \left[ \frac{b}{r^3}+2
\left(\frac{b}{r}-1\right) \frac{\Phi'}{r}
\right] \label{radialpWH}\,,\\
p_t(r)&=&\frac{1}{8\pi} \left(\frac{b}{r}-1\right)\Bigg[\Phi ''+
(\Phi')^2 + \frac{b'r+b-r}{2r(b-r)}\Phi'
\nonumber \\
&& +\frac{b'r-b}{2r^2(b-r)} \Bigg] \label{lateralpWH}\,.
\end{eqnarray}
in which $\rho(r)$ is the energy density, $p_r(r)$ is the radial
pressure, $p_t(r)$ is the pressure measured in the tangential
directions, orthogonal to the radial direction.

Assuming that the redshift function is finite $\forall r$, note
that the radial pressure is always positive at the throat, i.e,
$p_r=1/(8\pi r_0^2)$, contrary to the Morris-Thorne wormhole,
where a radial tension at the throat is needed to sustain the
wormhole. In addition to this, we also show below that $b'(r_0)>1$
at the throat, which implies a negative energy density at the
throat. This condition is another significant difference to the
Morris-Thorne wormhole, where the existence of negative energy
densities at the throat is not a necessary condition.

By taking the derivative with respect to the radial coordinate
$r$, of Eq. (\ref{radialpWH}), and eliminating $b'$ and $\Phi''$,
given in Eq. (\ref{rhoWH}) and Eq. (\ref{lateralpWH}),
respectively, we obtain the following equation
\begin{equation}
p_r'=-(\rho+p_r)\Phi '+\frac{2}{r}(p_t-p_r) \label{tauderivative}
\,.
\end{equation}
Equation (\ref{tauderivative}) is the relativistic Euler equation,
or the hydrostatic equation for equilibrium for the material
threading the hyperbolic spacetime tunnel.

We now have a system of three equations, namely, Eqs.
(\ref{rhoWH})-(\ref{lateralpWH}), with five unknown functions of
$r$, i.e., the stress-energy components, $\rho(r)$, $p_r(r)$ and
$p_t(r)$, and the metric fields, $b(r)$ and $\Phi(r)$. To
construct specific solutions, we may adopt several approaches, and
in this work we shall mainly use the strategy of considering a
specific equation of state given by $p_r=p_r(\rho)$, and
restricted choices for $b(r)$ or $\Phi(r)$. One may also impose a
specific form for the stress-energy components and through the
field equations and the equation of state, $p_r=p_r(\rho)$,
determine $b(r)$ and $\Phi(r)$. We will show below, through the
embedding analysis, that $b'(r_0)>1$, so that throughout this
paper we only consider the cases of a negative energy density at
the throat, $\rho|_{r_0}<0$, .


\subsection{Mathematics of embedding}\label{Sec:embedding}

The embedding diagrams are useful to represent the geometry of the
tunnelling solution and extract some useful information for the
choice of the shape function, $b(r)$ \cite{Misner}. Due to the
{\it pseudo}-spherically symmetric nature of the problem, without
a significant loss of generality consider a slice with the
specific value of $u=u_0$ which imposes $\sinh(u_0)=1$. The
respective line element, considering a fixed moment of time,
$t={\rm const}$, is given by
\begin{equation}
{\rm d}s^2=\frac{{\rm d}r^2}{b(r)/r-1}+r^2 \, {\rm d}v ^2\,.
\label{surface1}
\end{equation}
To visualize this slice, one embeds this metric into
three-dimensional Euclidean space \cite{Misner}, in which the
metric can be written in cylindrical coordinates, $(r,v,z)$, as
\begin{equation}
{\rm d}s^2={\rm d}z^2+{\rm d}r^2+r^2 \, {\rm d}v ^2 \,.
\end{equation}

\begin{figure}
\centering
\includegraphics[width=2.8in]{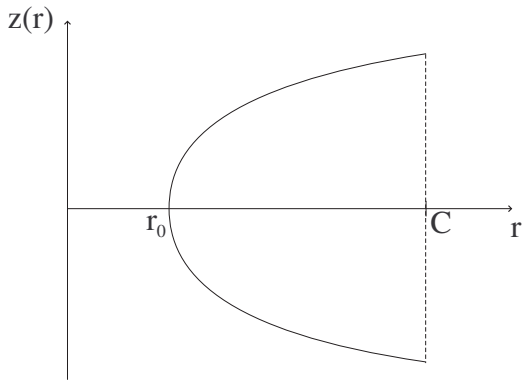}
\caption[Embedding diagrams of traversable wormholes]{The
embedding diagram of a two-dimensional section along the slice
($t={\rm const}$, $u=u_0 = \sinh^{-1}(1)$) of a tunnel in
spacetime.}\label{fig:embed2}
\end{figure}

Now, in the three-dimensional Euclidean space the embedded surface
has equation $z=z(r)$, and thus the metric of the surface can be
written as,
\begin{equation}
{\rm d}s^2=\left [1+\left( \frac{dz}{dr}\right)^2\right] {\rm
d}r^2+r^2 \, {\rm d}v ^2 \,. \label{surface2}
\end{equation}
Comparing Eq. (\ref{surface2}) with (\ref{surface1}), we have the
equation for the embedding surface, given by
\begin{equation}
\frac{dz}{dr}=\pm \left[\frac{2r-b(r)}{b(r)-r}\right]^{1/2}
\label{lift}\,.
\end{equation}
To be a solution of a spacetime tunnel, the geometry has a minimum
radius, $r=b(r)=r_{\rm 0}$, denoted as the throat, at which the
embedded surface is vertical, i.e., $dz/dr \rightarrow \infty$,
see Figure \ref{fig:embed2}. Note also that contrary to the
Morris-Thorne traversable wormhole, the shape function is
constrained in the present case. More specifically, taking into
account that $b(r)\geq r$, then the embedding surface (\ref{lift})
also imposes the condition $b(r)\leq 2r$. Thus, the shape function
is restricted to lie within the following range: $r\leq b(r) \leq
2r$.

In addition to this, to be a solution of a tunnel in spacetime,
one needs to impose that the throat flares out, as in Figure
\ref{fig:embed2}. Mathematically, this flaring-out condition
entails that the inverse of the embedding function $r(z)$, must
satisfy $d^2r/dz^2>0$ at or near the throat $r_{\rm 0}$.
Differentiating $dr/dz=\pm (r/b(r)-1)^{1/2}$ with respect to $z$,
we have
\begin{equation}
\frac{d^2r}{dz^2}=\frac{b'r-b}{2(2r-b)^2}>0   \label{flareout}\,.
\end{equation}
At the throat we verify that the shape function satisfies the
condition $b'(r_0)>1$, also contrary to its Morris-Thorne
counterpart. This condition plays a fundamental role in the
analysis of the violation of the energy conditions.

In particular, considering the NEC, i.e., $p_r+\rho \geq 0$, the
term $p_r+\rho$, taking into account the field equations
(\ref{rhoWH}) and (\ref{radialpWH}), is given by
\begin{equation}
\rho+p_r=\frac{1}{8\pi}\left[\frac{b-rb'}{r^3}
+2\left(\frac{b}{r}-1\right)\frac{\Phi'}{r}\right] \,.
\end{equation}
At the throat, we have
\begin{equation}
\left(\rho+p_r\right)|_{r_0}=\frac{1-b'(r_0)}{8\pi r_0^2} <0
 \,,
\end{equation}
which implies the important condition $b'(r_0)>1$ at the throat,
contrary to the Morris-Thorne counterpart. As mentioned above,
this implies that the energy density, at the throat, is always
negative for these exotic hyperbolic, tunneling geometries.


\section{Specific solutions}
\label{Sec:III}

In this Section, we find exact solutions by considering several
appropriate equations of state, $p_r=p_r(\rho)$, and by imposing
specific shape functions or specific redshift functions. To this
effect, an adequate shape function is the following case:
\begin{equation}
b(r)=r_0\left(\frac{r}{r_0}\right)^{\delta}
 \,. \label{specform}
\end{equation}
Taking the radial derivative, we have
\begin{equation}
b'(r)=\delta\left(\frac{r}{r_0}\right)^{\delta-1}
 \,,
\end{equation}
which at the throat implies $b'(r_0)=\delta>1$. From the condition
$b(r)<2r$, one has $r<2^{1/(\delta-1)}r_0$. Thus, we have the
following conditions:
\begin{eqnarray}
{\rm If} \qquad \delta \rightarrow \infty\,, \qquad {\rm then}
\qquad
r\rightarrow r_0 \\
{\rm If} \qquad \delta \rightarrow 1\,, \qquad {\rm then} \qquad
r\rightarrow \infty
 \,.
\end{eqnarray}
We verify that one may have an arbitrary large tunnel
by
imposing the condition $\delta \rightarrow 1$.


\subsection{Linear equation of state}

\subsubsection{Specific shape function}

Consider the linear equation of state $p_r(r)=\omega \rho(r)$, in
which wormhole solutions were extensively analyzed in Ref.
\cite{phantomWH}, and consequently denoted as ``phantom
wormhole''. Now, using the above-mentioned equation of state and
taking into account Eqs. (\ref{rhoWH})-(\ref{radialpWH}), one
deduces the following expression
\begin{equation}
\Phi'(r)=-\frac{b(r)+\omega b'(r)r}{2r\left[b(r)-r\right]}
 \,, \label{diffeq}
\end{equation}
which is the relationship governing what may be considered the
hyperbolic analogue of the ``phantom wormhole''. Note that the NEC
violation, and taking into account $\rho|_{r_0}<0$, imposes that
$\omega>-1$, contrary to its spherically symmetric counterpart.

By imposing a specific redshift function, one finds the general
solution for the shape function
\begin{equation}
b(r)=r^{-\frac{1}{\omega}}e^{-\frac{2}{\omega}\Phi}\;
\left[\frac{2}{\omega}\int
\Phi'r^{1+\frac{1}{\omega}}e^{\frac{2}{\omega}\Phi}\;dr
+r_0^{1+\frac{1}{\omega}}e^{\frac{2}{\omega}\Phi(r_0)} \right]\,.
\end{equation}

However, one also deduces an adequate solution by imposing the
shape function given by Eq. (\ref{specform}). Using relationship
(\ref{diffeq}), the redshift function is provided by
\begin{equation}
\Phi(r)=-\frac{1+\delta
\omega}{2(\alpha-1)}\ln\left[\left(\frac{r}{r_0}\right)^{\delta-1}
-1\right]+C_1\,,
\end{equation}
where $C_1$ is an integration constant. Now, to avoid an event
horizon at $r_0$, one needs to impose that $\omega=-1/\delta$, so
that the redshift function simplifies to $\Phi(r)=C_1$.

One may also further restrict the parameters by considering
certain traversability conditions in analogy to the Morris-Thorne
wormhole \cite{Morris:1988cz}. In particular, it is important that
an observer traversing through the hyperbolic tunnel should not be
ripped apart by enormous tidal forces. Thus, the tidal
traversability condition requires that the tidal accelerations
felt by the traveller should not exceed, for instance, the Earth's
gravitational acceleration, $g_\oplus$, which is translated by the
following inequalities
\begin{eqnarray}
2\left |\left (\frac{b}{r}-1 \right ) \left [\Phi ''+(\Phi ')^2+
\frac{b'r-b}{2r(r-b)}\Phi' \right] \right | &\leq & g_\oplus \,,
    \label{radialtidalconstraint}    \\
\left | \frac{\gamma ^2}{r^2} \left [v^2\left (\frac{b}{r}-b'
\right )+2(b-r)\Phi ' \right] \right |  &\leq & g_\oplus \,.
\label{lateraltidalconstraint}
\end{eqnarray}
We refer the reader to \cite{Morris:1988cz} for details in
deducing these relationships. The radial tidal constraint, Eq.
(\ref{radialtidalconstraint}), constrains the redshift function,
and the lateral tidal constraint, Eq.
(\ref{lateraltidalconstraint}), constrains the velocity with which
observers traverse the tunnel. These inequalities are
particularly simple at the throat, $r_0$,
\begin{eqnarray}
|\Phi '(r_0)| &\leq & \frac{g_{\oplus}\,r_0}{(b'-1)} \,,
      \label{radialtidalconstraint2}    \\
\gamma^2 v^2 &\leq & \frac{g_{\oplus}\,r_0^2}{(b'-1)} \,.
 \label{lateraltidalconstraint2}
\end{eqnarray}

In particular, considering non-relativistic velocities, i.e.,
$v\ll 1$ and $\gamma \sim 1$, using the shape function given by
(\ref{specform}) and $\Phi(r)=C_1$, one readily verifies that the
radial tidal constraint is satisfied. The lateral tidal constraint
provides the condition $v\leq g_{\oplus}r_0^2/(\delta-1)$, which
essentially depends on the values taken for the parameter
$\delta$. In alternative, by specifying the traversal velocity, we
find a restriction placed on $\delta$ given by $\delta\leq
1+g_{\oplus}r_0^2/v$.

\subsubsection{Specific redshift function}

It is also adequate to find solutions by considering a specific
redshift function. Consider the redshift function given by
$\Phi(r)=r_0/r$. Using expression (\ref{diffeq}), one deduces the
shape function given by
\begin{eqnarray}
b(r)&=&\frac{4r_0^2}{\omega}\left(-\frac{2r_0}{\omega r
}\right)^{\frac{1-\omega}{\omega}}e^{-\frac{2r_0}{\omega r}} {\cal
F} \left(\frac{\omega-1}{\omega},-\frac{2r_0}{\omega r }\right)
    \nonumber \\
&&-2r_0+e^{-\frac{2r_0}{\omega r}}r^{-\frac{1}{\omega}}C_2
 \,,
\end{eqnarray}
where $C_2$ is an integration constant and, for notational
simplicity, the function ${\cal F}$ is defined as
\begin{equation}
{\cal F} \left(\frac{\omega-1}{\omega},-\frac{2r_0}{\omega r
}\right)=\Gamma \left(\frac{\omega-1}{\omega}\right)-\Gamma
\left(\frac{\omega-1}{\omega},-\frac{2r_0}{\omega r }\right)\,,
\end{equation}
and $\Gamma(x)$ and $\Gamma(x,z)$ are the Gamma and the incomplete
Gamma functions, respectively. Note that the constant of
integration $C_2$ may be found by taking into account the
condition $b(r_0)=r_0$, and is given by the following relationship
\begin{equation}
C_2=r_0^{1+\frac{1}{\omega}}\left[3e^{\frac{2}{\omega}}
+2\left(-\frac{2}{\omega}\right) ^{\frac{1}{\omega}} {\cal F}
\left(\frac{\omega-1}{\omega},-\frac{2r_0}{\omega r
}\right)\right]\,.
\end{equation}


\subsection{``Generalized Chaplygin gas'' equation of state}

In cosmology, the equation of state representing the generalized
Chaplygin gas (GCG) is given by $p_{ch}=-A/\rho_{ch}^{\alpha}$,
where $A$ and $\alpha$ are positive constants, and the latter lies
in the range $0< \alpha \leq 1$ \cite{Kamen,GCGbrane2}. The
particular case of $\alpha=1$ corresponds to the Chaplygin gas. An
attractive feature of this model is that at early times the energy
density behaves as matter, $\rho_{ch}\sim a^{-3}$, where $a$ is
the scale factor, and as a cosmological constant at a later stage,
i.e., $\rho_{ch}={\rm const}$. In a cosmological context, at a
late stage dominated by an accelerated expansion of the Universe,
the cosmological constant may be given by $8\pi A^{1/(1+\alpha)}$.
This dual behavior is responsible for the interpretation that the
GCG model is a candidate of a unified model of dark matter and
dark energy \cite{BBS}.

It was noted in Ref. \cite{BertPar} that the GCG equation of state
is that of a polytropic gas with a negative polytropic index, and
thus suggested that one could analyze astrophysical implications
of the model. In this context, the construction of traversable
wormholes, possibly arising from a density fluctuation in the GCG
cosmological background was explored in \cite{ChapWH}. These
latter solutions were denoted Chaplygin wormholes. As in Ref.
\cite{ChapWH}, it was considered that the pressure in the GCG
equation of state is a radial pressure, and the tangential
pressure can be determined from the Einstein equations, in
particular, Eq. (\ref{lateralpWH}).

To compare to the spherically symmetric Chaplygin wormholes,
we construct their pseudo-spherically symmetric duplicates. Thus,
taking into account the GCG equation of state in the form
$p_r=-A/\rho^{\alpha}$, and using Eqs.
(\ref{rhoWH})-(\ref{radialpWH}), we have the following condition
\begin{equation}
\Phi'(r)=\left[A(8\pi)^{1+\alpha}\;\frac{r^{2\alpha +1}
}{2(b')^{\alpha}}-\frac{b}{2r^2}\right]\Big/\left(\frac{b}{r}
-1\right) \,.
            \label{EOScondition}
\end{equation}
Solutions of the metric (\ref{metricwormhole}), satisfying Eq.
(\ref{EOScondition}) may be considered as the analogues of
``Chaplygin wormholes'' analyzed in \cite{ChapWH}.

As shown above, to be a spacetime tunnel, the condition
$b'(r_0)>1$ is imposed. Now, using the GCG equation of state,
evaluated at the throat, and taking into account Eq.
(\ref{radialpWH}), we verify that the energy density at $r_0$ is
given by $\rho(r_0)=-\left(8\pi r_0^2 A\right)^{1/\alpha}$.
Finally, using Eq. (\ref{rhoWH}), and the condition $b'(r_0)>1$,
we verify that for these solutions, the following
condition is imposed $A> \left(8\pi r_0^2\right)^{-(1+\alpha)}$,
contrary to its spherically symmetric counterpart \cite{ChapWH}.

We consider next a specific example, by taking into account the
shape function given by Eq. (\ref{specform}). Thus, the
differential equation (\ref{EOScondition}) provides the following
redshift function
\begin{eqnarray}
\Phi(r)&=&\frac{1}{2}\Bigg\{\left(\frac{r}{r_0}\right)^\delta r_0
\left[\sum_{k=0}^{-1}\,\left(\frac{1}{1+k}\right)
\left(\frac{r}{r_0}\right)^{k(\delta-1)} \right]
  \nonumber   \\
&&
-A(8\pi)^{1+\alpha}\left[\delta\left(\frac{r}{r_0}\right)^{\delta-1}
\right]
r^{2\alpha+3} \times
  \nonumber  \\
&&\times {\rm
LerchPhi}\left[\left(\frac{r}{r_0}\right)^{\delta-1},1,
\frac{3\alpha+2-\alpha\delta}{\delta-1}\right]
    \nonumber  \\
&&-r\ln\left[\left(\frac{r}{r_0}\right)^{\delta-1}-1\right]
 \Bigg\}/\left[(\delta-1)r\right]+C_3\,,
\end{eqnarray}
where $C_3$ is a constant of integration; and LerchPhi is the
general Lerch Phi function, defined as ${\rm
LerchPhi}(z,a,v)=\sum_{n=0}^{\infty}\;z^n/(v+n)^a$.

It is interesting to consider the specific case of $\delta=2$ and
$\alpha=1$ (this latter value corresponds to the Chaplygin gas
equation of state), which yields the following simpler solution
\begin{eqnarray}
\Phi(r)&=&8\pi^2Ar_0^3
r\left(2+\frac{r}{r_0}\right)
   \nonumber   \\
&&+\frac{1}{2}\left(8\pi
Ar_0^2-1\right)\ln\left(\frac{r}{r_0}-1\right)+C_3\,,
\end{eqnarray}
Note the existence of an event horizon at the throat, rendering the hyperbolic tunnel non-traversable. Thus, to avoid this we impose the
condition $A=1/(8\pi r_0^2)$, so that the above solution
simplifies to
\begin{equation}
\Phi(r)=8\pi^2Ar_0^3 r\left(2+\frac{r}{r_0}\right)+C_3\,,
\end{equation}
yielding a traversable geometry.


\subsection{Van der Waals equation of state}

Another case that lends itself to our analysis, is that of the van
der Waals quintessence equation of state, which seems to provide a
solution to the puzzle of dark energy, without the presence of
exotic fluids or modifications of the Friedmann equations. In Ref.
\cite{vdWWH}, the construction of inhomogeneous compact spheres
supported by a van der Waals equation of state was explored. These
relativistic stellar configurations were denoted as {\it van der
Waals quintessence stars}. Despite the fact that, in a
cosmological context, the van der Waals fluid is considered
homogeneous, inhomogeneities may arise through gravitational
instabilities. Thus, these solutions may possibly originate from
density fluctuations in the cosmological background. Exact
solutions were found, and their respective characteristics and
physical properties were further explored in Ref. \cite{vdWWH}.

The van der Waals equation of state is given by
\begin{equation}
p=\frac{\gamma \rho}{1-\beta\rho}-\alpha\rho^2\,,
\end{equation}
where $\rho$ is the energy density and $p$ the pressure of the VDW
fluid. The accelerated and decelerated periods depend on the
parameters, $\alpha$, $\beta$ and $\gamma$ of the equation of
state, and in the limiting case $\alpha,\beta \rightarrow 0$, one
recovers the dark energy equation of state, with $\gamma=p/\rho
<-1/3$ \cite{Capo3}.

As in the previous exact solutions, we may also consider the
counterpart of the static and spherically symmetric van der Waals
wormhole. Thus, taking into account the specific shape function
given by Eq. (\ref{specform}), with the parameter $\delta=2$, and
considering that $p$ is a radial pressure (see \cite{vdWWH} for
details), we have the following solution
\begin{eqnarray}
\Phi(r)&=&-\frac{1}{4\pi
r_0^2\overline{\beta}}\Bigg[(\overline{\alpha}
\overline{\beta}+16\pi^2r_0^4\gamma)\ln\left(1-\frac{r_0}{r}\right)
  \nonumber \\
&&+4\pi r_0^2\beta \gamma\ln\left(1+\frac{4\pi
r_0r}{\beta}\right)\Bigg]+C_4
 \,,
\end{eqnarray}
where $C_4$ is a constant of integration, and we have considered
the following definitions for notational simplicity:
\begin{equation}
\overline{\alpha}=\alpha+2\pi r_0^2\;, \qquad
\overline{\beta}=\beta+4\pi r_0^2 \,.
\end{equation}

Note the presence of an event horizon, due to the term
$\ln(1-r_0/r)$. Thus, to avoid this, we impose the following
condition
\begin{equation}
\gamma=-\frac{\overline{\alpha}\overline{\beta}}{16\pi^2r_0^4}=-
\frac{(\alpha+2\pi r_0^2)(\beta+4\pi r_0^2)}{16\pi^2r_0^4} \,.
\end{equation}
so the redshift function reduces to
\begin{equation}
\Phi(r)=\frac{(\alpha+2\pi
r_0^2)\beta}{16\pi^2r_0^4}\ln\left(1+\frac{4\pi
r_0r}{\beta}\right)+C_4
 \,.  \label{vdWredshift}
\end{equation}

Using the simplified relationship (\ref{vdWredshift}), one may
find a restriction from the lateral tidal acceleration constraint
(\ref{lateraltidalconstraint2}), provided by
\begin{equation}
|\overline{\alpha}| \leq 4\pi r_0^4 g_{\oplus}
|\overline{\beta}/\beta|  \,.
\end{equation}


\section{Summary, discussion and future outlook}
\label{Sec:conclusion}

Traversable wormholes possess a peculiar property, namely ``exotic
matter'', involving a stress-energy tensor that violates the null
energy condition. In this work, we have constructed exact
solutions of static and {\it pseudo}-spherically symmetric
spacetime tunnels by adding exotic matter to a vacuum solution
referred to as a degenerate solution of class $A$. The usual
2-dimensional spheres are replaced by pseudo-spheres, which are
still surfaces of revolution around an axis, but now consist of a
negative and constant curvature. The physical properties and
characteristics of these intriguing, hyperbolic solutions were
further explored, and through the mathematics of embedding it was
shown that particular constraints are placed on the shape
function, that differ radically from the Morris-Thorne wormhole.
In particular, it was shown that the energy density is always
negative and the radial pressure is positive, at the throat,
contrary to the Morris-Thorne counterpart. Specific solutions were
also presented by considering several different equations of
state, and by imposing restricted choices for the shape function
or the redshift function.

Standard definitions of a wormhole usually require either
asymptotic flatness with a localized bridge to another
asymptotically flat region, or the ability to identify a compact
region that can usefully be thought of as a throat. The geometries
considered in the current article also exhibit asymptotic flatness
and share the possibility of having localized tunnels, with
throats, bridging different spacetime regions. However the spatial
hypersurfaces of the asymptotic flat limit of the hyperbolic
spacetime under consideration here is peculiar in that it is not
euclidean, but rather pseudo-euclidean (it is a 3-dimensional
Lorentzian space). Hence, rather than denote the exotic
geometric objects analyzed in this work as wormholes, we prefer the term
hyperbolic tunnels. They should indeed be understood as a
theoretician's probe of the foundations of general relativity, on
the same grounds as in the case of the Morris-Thorne wormhole.

\section*{Acknowledgments}

The authors are thankful to Reza Tavakol for helpful comments, and
gratefully acknowledge the grants PTDC/FIS/102742/2008 and
CERN/FP/109381/2009 from FCT.




\begin{thebibliography}{99}

\bibitem{Morris:1988cz}
  M.~S.~Morris and K.~S.~Thorne,
  ``Wormholes in space-time and their use for interstellar travel:
  A tool for teaching general relativity,''
  Am.\ J.\ Phys.\  {\bf 56}, 395 (1988).

\bibitem{Visser}
  M.~Visser,
  {\it Loretzian wormholes: from Einstein to Hawking}
  AIP Press (1995).

\bibitem{EGB1}
B. Bhawal and S. Kar, ``Lorentzian wormholes in
Einstein-Gauss-Bonnet theory,'' Phys. Rev. D {\bf 46}, 2464-2468
(1992).

\bibitem{EGB2}
G. Dotti, J. Oliva, and R. Troncoso, ``Static wormhole solution
for higher-dimensional gravity in vacuum,'' Phys. Rev. D {\bf 75},
024002 (2007) [arXiv:hep-th/0607062].

\bibitem{braneWH1}
L. A. Anchordoqui and S. E. P Bergliaffa, ``Wormhole surgery and
cosmology on the brane: The world is not enough,'' Phys. Rev. D
{\bf 62}, 067502 (2000) [arXiv:gr-qc/0001019];
%
K. A. Bronnikov and S.-W. Kim, ``Possible wormholes in a brane
world,'' Phys. Rev. D {\bf 67}, 064027 (2003)
[arXiv:gr-qc/0212112];
%
M. La Camera, ``Wormhole solutions in the Randall-Sundrum
scenario,'' Phys. Lett. {\bf B573}, 27-32 (2003)
[arXiv:gr-qc/0306017].

\bibitem{braneWH2}
F.~S.~N.~Lobo,
  ``General class of braneworld wormholes,''
  Phys.\ Rev.\ {\bf D75}, 064027 (2007)
  [arXiv:gr-qc/0701133].

\bibitem{Nandi:1997en}
K.~K.~Nandi, B.~Bhattacharjee, S.~M.~K.~Alam and J.~Evans,
  ``Brans-Dicke wormholes in the Jordan and Einstein frames,''
  Phys.\ Rev.\  D {\bf 57}, 823 (1998);
  %
  F.~S.~N.~Lobo and M.~A.~Oliveira,
  Phys.\ Rev.\  D {\bf 81}, 067501 (2010);
%
F.~S.~N.~Lobo and M.~A.~Oliveira,
  Phys.\ Rev.\  D {\bf 80}, 104012 (2009).

\bibitem{Garattini:2007ff}
R.~Garattini and F.~S.~N.~Lobo,
  ``Self sustained phantom wormholes in semi-classical gravity,''
  Class.\ Quant.\ Grav.\  {\bf 24}, 2401 (2007)
  [arXiv:gr-qc/0701020];
%
R.~Garattini and F.~S.~N.~Lobo,
  ``Self-sustained traversable wormholes in noncommutative geometry,''
  Phys.\ Lett.\  B {\bf 671}, 146 (2009)
  [arXiv:0811.0919 [gr-qc]].


\bibitem{Boehmer:2007rm}
  C.~G.~Boehmer, T.~Harko and F.~S.~N.~Lobo,
  ``Conformally symmetric traversable wormholes,''
  Phys.\ Rev.\  D {\bf 76}, 084014 (2007)
  [arXiv:0708.1537 [gr-qc]];
%
C.~G.~Boehmer, T.~Harko and F.~S.~N.~Lobo,
  ``Wormhole geometries with conformal motions,''
  Class.\ Quant.\ Grav.\  {\bf 25}, 075016 (2008)
  [arXiv:0711.2424 [gr-qc]].



\bibitem {phantomWH}
S.~Sushkov, ``Wormholes supported by a phantom energy,'' Phys.
Rev. D {\bf 71}, 043520 (2005) [arXiv:gr-qc/0502084];
%
F.~S.~N.~Lobo,
  ``Phantom energy traversable wormholes,''
  Phys.\ Rev.\ {\bf D71}, 084011 (2005)
  [arXiv:gr-qc/0502099];
%
  F.~S.~N.~Lobo,
  ``Stability of phantom wormholes,''
  Phys.\ Rev.\ {\bf D71}, 124022 (2005)
  [arXiv:gr-qc/0506001].

\bibitem{ChapWH}
F.~S.~N.~Lobo,
  ``Chaplygin traversable wormholes,''
  Phys.\ Rev.\ {\bf D73}, 064028 (2006)
  [arXiv:gr-qc/0511003].

\bibitem{vdWWH}
F.~S.~N.~Lobo,
  ``Van der Waals quintessence stars,''
  Phys.\ Rev.\  D {\bf 75}, 024023 (2007)
  [arXiv:gr-qc/0610118].

\bibitem{Weylgrav}
F.~S.~N.~Lobo,
  ``General class of wormhole geometries in conformal Weyl gravity,''
  Class.\ Quant.\ Grav.\  {\bf 25}, 175006 (2008)
  [arXiv:0801.4401 [gr-qc]].



\bibitem{Lemos:2003jb}
  J.~P.~S.~Lemos, F.~S.~N.~Lobo and S.~Quinet de Oliveira,
  ``Morris-Thorne wormholes with a cosmological constant,''
  Phys.\ Rev.\  D {\bf 68}, 064004 (2003)
  [arXiv:gr-qc/0302049].

\bibitem{Lobo:2007zb}
  F.~S.~N.~Lobo,
  ``Exotic solutions in General Relativity: Traversable wormholes and 'warp
  drive' spacetimes,''
  arXiv:0710.4474 [gr-qc].

\bibitem{Stephani:2003tm}
  H.~Stephani, D.~Kramer, M.~A.~H.~MacCallum, C.~Hoenselaers and E.~Herlt,
{\it  Cambridge, UK: Univ. Pr. (2003) 701 P}

\bibitem{Ehlers & Kundt 1962} J. Ehlers and W. Kundt, "Exact Solutions of the gravitational Field Equations", in Gravitation: an introduction to current research,
ed. L. Witten, pp49 (Wiley, New York and London, 1962)

\bibitem{Levi-Civita 1917a} T. Levi-Civita, Ren. Acc. Lincei 27, 183 (1017).

\bibitem{Weyl:1917gp}
  H.~Weyl,
  Annalen Phys.\  {\bf 54} (1917) 117.

\bibitem{Bonnor & Martins 1991} W.~B.~Bonnor and M-~A.~P.~Martins, Classical and Quantum Gravity, 8, 727 (1991).
\bibitem{Martins 1996} M. A. P. Martins, Gen. Rel. Grav. 28 (1996) 1309.

\bibitem{Anchordoqui:1995wa}
  L.~A.~Anchordoqui, J.~D.~Edelstein, C.~Nunez and G.~S.~Birman,
  arXiv:gr-qc/9509018.


\bibitem{MimLobo1}
  J.~P. Mimoso and F.~S.~N.~Lobo,
  ``Vacuum solutions with pseudo-spherical symmetry,'' in
  preparation.

\bibitem{MimLobo2}
  J.~P. Mimoso and F.~S.~N.~Lobo, in preparation.

\bibitem{Visser89}
M. Visser, ``Traversable wormholes: Some simple examples'', Phys.
Rev. D {\bf 39}, 3182 (1989).

\bibitem{GDiaz}
P. F. Gonz\'alez-D\'iaz, ``Ringholes and closed timelike curves'',
Phys. Rev. D {\bf 54}, 6122 (1996).

\bibitem{LL}
  J.~P.~S.~Lemos and F.~S.~N.~Lobo,
  ``Plane symmetric traversable wormholes in an anti-de Sitter
  background,''
  Phys.\ Rev.\  D {\bf 69}, 104007 (2004)
  [arXiv:gr-qc/0402099].

\bibitem{LL2}
  J.~P.~S.~Lemos and F.~S.~N.~Lobo,
  ``Plane symmetric thin-shell wormholes: Solutions and stability,''
  Phys.\ Rev.\  D {\bf 78}, 044030 (2008)
  [arXiv:0806.4459 [gr-qc]].


\bibitem{dynamicWH}
D.~Hochberg and M.~Visser,
  ``The null energy condition in dynamic wormholes,''
  Phys.\ Rev.\ Lett.\  {\bf 81}, 746 (1998)
  [arXiv:gr-qc/9802048];
%
  D.~Hochberg and M.~Visser,
  ``Dynamic wormholes, anti-trapped surfaces, and energy conditions,''
  Phys.\ Rev.\  D {\bf 58}, 044021 (1998)
  [arXiv:gr-qc/9802046];
%
S.~Kar,
  ``Evolving wormholes and the weak energy condition,''
  Phys.\ Rev.\  D {\bf 49}, 862 (1994);
%
 S.~Kar and D.~Sahdev,
  ``Evolving Lorentzian wormholes,''
  Phys.\ Rev.\  D {\bf 53}, 722 (1996)
  [arXiv:gr-qc/9506094];
%
S.~W.~Kim,
  ``The Cosmological model with traversable wormhole,''
  Phys.\ Rev.\  D {\bf 53}, 6889 (1996);
%
A.~V.~B.~Arellano and F.~S.~N.~Lobo,
  ``Evolving wormhole geometries within nonlinear electrodynamics,''
  Class.\ Quant.\ Grav.\  {\bf 23}, 5811 (2006)
  [arXiv:gr-qc/0608003].

\bibitem{mty}
M. S. Morris, K. S. Thorne and U. Yurtsever, ``Wormholes, Time
Machines and the Weak Energy Condition,'' Phy. Rev. Lett. {\bf
61}, 1446 (1988).

\bibitem{Misner}
C. W. Misner, K. S. Thorne and J. A. Wheeler, {\it Gravitation}
(W. H. Freeman and Company, San Francisco, 1973).

\bibitem{Kamen}
A. Y. Kamenshchik, U. Moschella and V. Pasquier, ``An alternative
to quintessence,'' Phys. Lett. B {\bf 511}, 265 (2001),
[arXiv:gr-qc/0103004].

\bibitem{GCGbrane2}
M. C. Bento, O. Bertolami and A. A. Sen, ``Generalized Chaplygin
Gas, Accelerated Expansion and Dark Energy-Matter Unification,''
Phys. Rev. D {\bf 66}, 043507 (2002) [arXiv:gr-qc/0202064].

\bibitem{BBS}
M. C. Bento, O. Bertolami, A. A. Sen, ``The revival of the unified
dark energy-dark matter model?'' Phys.Rev. D {\bf 70}, 083519
(2004) [arXiv:astro-ph/0407239].

\bibitem{BertPar}
O. Bertolami and J. Paramos, ``The Chaplygin dark star,'' Phys.
Rev. D {\bf 72} 123512 (2005) [arXiv:astro-ph/0509547].

\bibitem{Capo3}
S. Capozziello, V. F. Cardone, S. Carloni, D. De Martino, M.
Falanga, A. Troisi and M. Bruni, ``Constraining van der Waals
quintessence by observations,'' JCAP 0504, 005 (2005),
[arXiv:astro-ph/0410503].

\bibitem{voids}
P.~J.~E. Peebles, Ap. J, 557, 495 (2001);
  V.~G.~Gurzadyan and A.~A.~Kocharyan,
  Astron.\ Astrophys.\  {\bf 493} (2009) L61
  [arXiv:0807.1239 [astro-ph]];
  T.~Clifton, P.~G.~Ferreira and K.~Land,
Phys.\ Rev.\ Lett.\  {\bf 101} (2008) 131302
  [arXiv:0807.1443 [astro-ph]].


\end{thebibliography}
\end{document}